FRI-2G.303-1-CCT1-05

# EXPERIMENTAL EVALUATION OF THE PHP'S CURL LIBRARY PERFORMANCE[5]

**Yordan Kalmukov, PhD**
Department of Computer Systems and Technologies,
University of Ruse
E-mail: jkalmukov@uni-ruse.bg

*Abstract: cURL (libcurl) is a popular and widely used library distributed with the php interpreter. It allows php applications to connect to and communicate with external resources (servers) by using wide variety of communication protocols. In most cases it is the preferred way of consuming external REST web services. Programmers usually use it for granted without even thinking of any performance issues. During an experimental analysis of the Hadoop's WebHDFS API throughput, it has been noted that read (download) speed from WebHDFS reduces with increasing the file size. However, this issue does not happen when writing to WebHDFS. Since the communication between the php application and the WebHDFS API is handled by the php's cURL library, then the cause of the download speed decrease could be either the cURL library itself or the API.*

*This paper presents a series of experimental analyses aiming to determine the cause of the download speed decrease in previous experiments – whether it is the WebHDFS API or the php's cURL library. Both parties are tested in multiple ways separately and independently of each other. Results clearly prove (in two different ways) that the cause of the download speed decrease is the php's cURL library itself, not the consumed API.*

## INTRODUCTION

The cURL project (https://curl.se) is an important milestone in the IT industry. Started in 1998, its aim is to provide free open source standalone tools and libraries for various languages that allow programmers and applications to connect to and communicate with external resources/servers by using wide variety of (virtually all) communication protocols. Unix, Linux and MacOS integrate it in their operating systems by default. A cURL library is available for all popular programming languages, distributed together with the compiler/interpreter. The cURL libraries are used not only by server/desktop applications, but by any other Internet-connected devices such as cars, TV sets, printers, routers, audio equipment, phones, tablets, smart home appliances – fridges, air conditioners, toasters, irons, lights and other IoT devices. Just a small off-topic deviation here: In my personal opinion it is not a good idea that all electronic devices should be connected to the Internet, since most of the small, energy-efficient IoT gadgets have not enough processing power to implement and support reliable security protocols, so they are quite easily hacked and zombified to form a large distributed bot-nets that could be subsequently used to perform distributed denial of service (DDoS) attacks.

When building distributed web applications and systems, cURL is the preferred way of communication between the separate web services, especially for REST-based services. Distributed computing is important not just for solving hard, computationally-intensive task, but it is crucial for heavy-loaded daily-life services such as Google, Facebook and etc., which have millions or billions of users. As the number of processors and the amount of memory are physically limited per single computer system, it is impossible for a single system to handle heavy load generated by millions of users. Since further vertical scaling is impossible, then the horizontal scaling by adding more machines remains the only feasible option. One of the most popular scalable systems/frameworks for distributed computing, big data processing and storage is the Apache Hadoop (https://hadoop.apache.org). It is an entire ecosystem of distributed applications that can run on

---

thousands of commodity servers and reliably store and process petabytes of data. Apache Spark (https://spark.apache.org) is just a single tool from this ecosystem that could be successfully used to process and analyze enormous amounts of batch data or streaming data in real time. It could be (and is) used in many applications, for example: document and text processing (Kalmukov, Y., 2020), (Kalmukov, Y., 2022); network traffic analysis (Aggarwal, S., 2023); data transformation, augmentation and enrichment (Aggarwal, S., 2023); profiling users (Aggarwal, S., 2023), (ProjectPro Team, 2023); and many others, in various industries such as finance, healthcare, media, e-commerce, retail, travel, gaming industries, government and corporate data processing and others (ProjectPro Team, 2023).

In order to use the big data processing and analytical tools of the Hadoop ecosystem, a company (or any organization in general) should first share its data with the Hadoop cluster. Together with a colleague, we have proposed an architecture for integration of a company's heterogeneous data to a remote Hadoop cluster (Kalmukov, Y., & Marinov, M., 2022). It allows storing and processing both streaming and batch data. In case of large amounts of batch data, the data files should be stored in the Hadoop Distributed File System (HDFS) first, then passed to Spark or other tools. There are several ways of saving data to HDFS from within the cluster, but from outside the most preferable way is the remote access through the WebHDFS API. It allows third-party applications to connect to remote HDFS file system and write/read files to/from it. Since WebHDFS is a REST-based API, the most common way of accessing it should be by using the cURL library from the chosen programming language. As we consider big data processing and analysis, the data files that should be stored in HDFS are supposed to be very large – in hundreds of megabytes or gigabytes. So the performance and the throughput of the cURL library really matters and could highly impact the processing time.

This paper presents a series of experimental analyses aiming to determine the php's cURL library performance and throughput. Both the WebHDFS API and the php's cURL library are tested in multiple ways separately and independently of each other.

**EXPOSITION**

**Motivation of performing experimental evaluation of the cURL library**

In previous experimental analysis of the WebHDFS API throughput (Kalmukov, Y., & Marinov, M., 2023), I noticed that read (download) speed from WebHDFS reduces with increasing the file size. Just to mention that the analysis was related to large amounts of data, i.e. files in hundreds of megabytes or gigabytes. The experimental application is built in PHP that is somehow atypical programming language for Hadoop, but since WebHDFS is a REST API, it could be accessed from any programming language or technology. The communication with the API is done through the PHP's cURL library. Results also show another unexpected issue – download speed decreases with the increase of the file size, but the upload (writing) speed remains the same and very high. Common sense suggests that if there is a decrease in transfer speed, it rather should be in the writing speed as files are replicated on multiple nodes. But upload speed is constant while download speed highly decreases with increasing the file size.

Since the communication between the PHP application and the WebHDFS API is handled by the PHP's cURL library, then the cause of the download speed decrease could be either the cURL library itself or the API. That motives me to do some further experiments and to test both the client and the API independently, in order to determine the guilty part.

**Experimental description and architecture**

As already mentioned, the experimental application is built in PHP. The communication with the WebHDFS API is done through the PHP's cURL library. The "PHP-Hadoop-HDFS" library does not perform any data processing at all, but just composes the necessary HTTP requests to access the WebHDFS API. The API could be accessed without PHP-Hadoop-HDFS library, but it facilitates the access, since the library frees the programmer from having to know the WebHDFS API itself. The architecture of the experimental system is presented on figure 1.

The Hadoop cluster consist of 10 rack-mounted servers - 1 name node (2x Intel Xeon Silver 4110, 32 threads / 64 GB RAM) and 9 data nodes (1x Intel Xeon E-2124, 4 cores / 16 GB RAM), connected through a 24-port gigabit switch, supporting 1 gbit/s per port. For all experiments, the client application runs on the same laptop computer – Intel i7 (4 cores) / 12 GB RAM, Windows 10 v1607, PHP 7.3.23, cURL library v7.70.

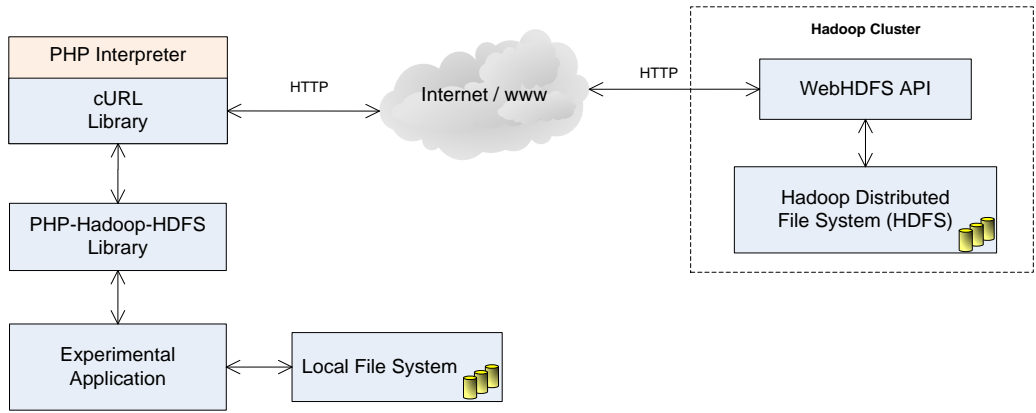

Fig. 1. Architecture of the original system for experimental study of WebHDFS API.

The WebHDFS API is also tested in alternative way by using the built-in cURL command in MacOS and an external stand-alone cURL application for Windows. Results show that files are downloading (read) from HDFS with very high speed almost reaching the maximum throughput capacity of the relevant type of network, regardless of the file size.

So, the causer of the read speed decrease in my previous experiments is, with a high degree of probability, the cURL library, distributed with the PHP interpreter. Thus, it should be separately and independently tested. To do so, the same data files are used, but they are uploaded to the file system of Apache HTTP server, rather than the distributed HDFS.

The revised architecture, for testing the cURL library only, is presented on figure 2. The "PHP-Hadoop-HDFS" library is removed since the cURL will perform plain HTTP GET requests, with no parameters, directly to the file resources. The Hadoop cluster is replaced with a single web server (Intel i5-4570 @ 3.20GHz, 4 cores, 16 GB RAM), running Apache HTTP Server version 2.4.41 on Windows 10 22H2. The server hardware is not of high importance since it is the same for all experiments and is powerful enough to handle very high speed communication. The client experimental application runs on the same laptop computer (Intel i7-7500U, 12 GB RAM, Windows 10 v1607, PHP 7.3.23, cURL library v7.70.) for all experiments in all types of networks. The client and the server are connected through a 1 gbit/s wired router. Experiments are performed on 1 gbit/s and 100 mbit/s connectivity.

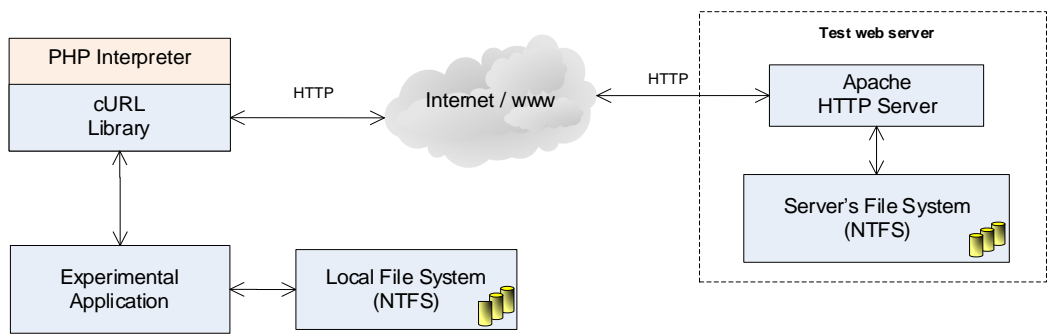

Fig. 2. Revised architecture of the system for experimental study of php's cURL performance.

For the current experiments, I use the same data files as in our previous experiments for studying the WebHDFS API throughput – small to medium (from 10 to 100 MB, with a step of 10MB), medium (from 100 to 300 MB, with a step of 50MB) and large (500 do 1500MB, with a step of 500MB).

**Results and discussion**

Results from the experimental study of the php's cURL library only also show that download (read) speed rapidly decreases with the increase of the files size, especially for large files. Results for small to medium files are shown on figure 3. Figure 4 presents results for medium file sizes, and figure 5 – for large files up to 1.5 GB.

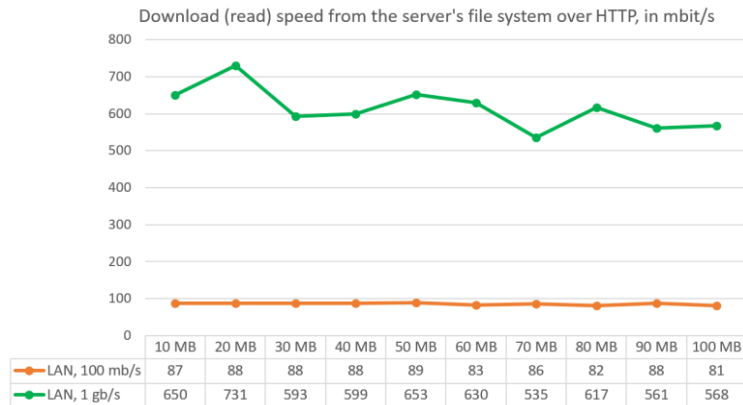

Fig. 3. Download (read) speed from the server's file system over HTTP, for files 10 to 100 MB.

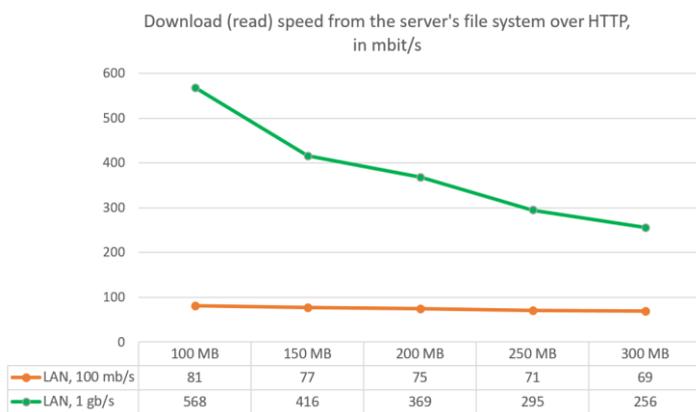

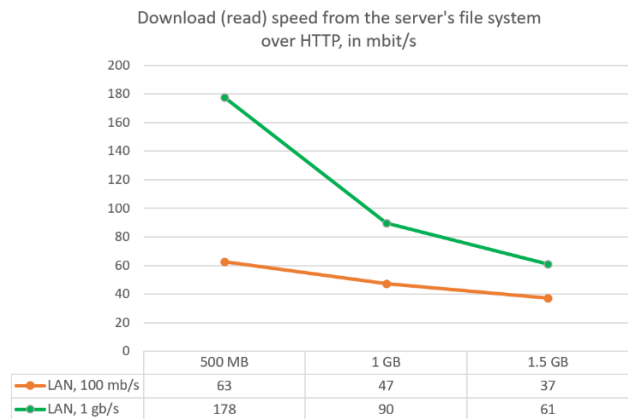

Fig. 4. Download (read) speed from the server's file system over HTTP, for files 100 to 300 MB.

Fig. 5. Download (read) speed from the server's file system over HTTP, for files 500 to 1500 MB.

It is obvious that the download speed rapidly decrease with increasing the files size. However it is interesting to directly compare these lines to the ones obtained during the experimental evaluation of the WebHDFS API. Thus, figures 6 and 7 present such a comparison for the medium and large files.

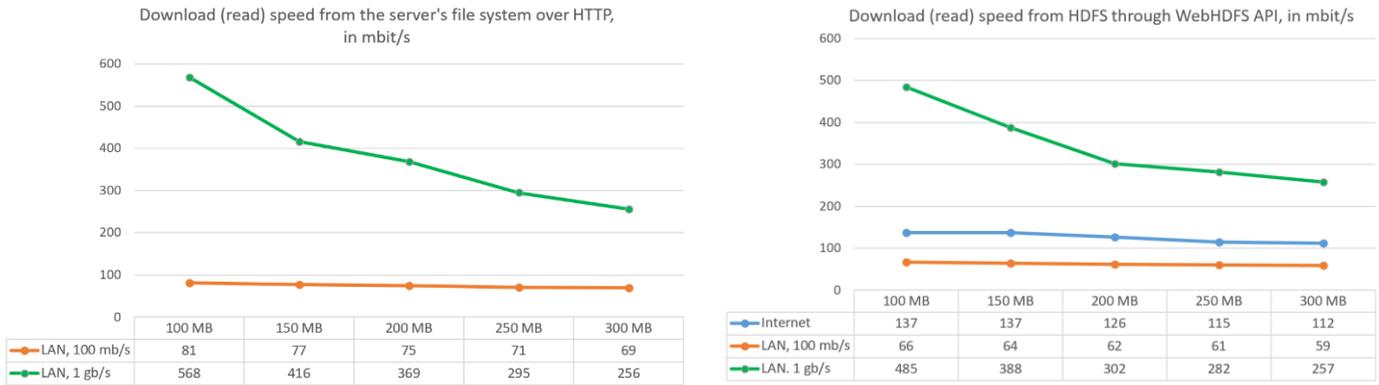

a) Download speed from the server's file system over HTTP    b) Download speed from HDFS through WebHDFS API

Fig. 6. Download (read) speed from the server's file system over HTTP
and from HDFS through WebHDFS API, for files from 100 to 300 MB

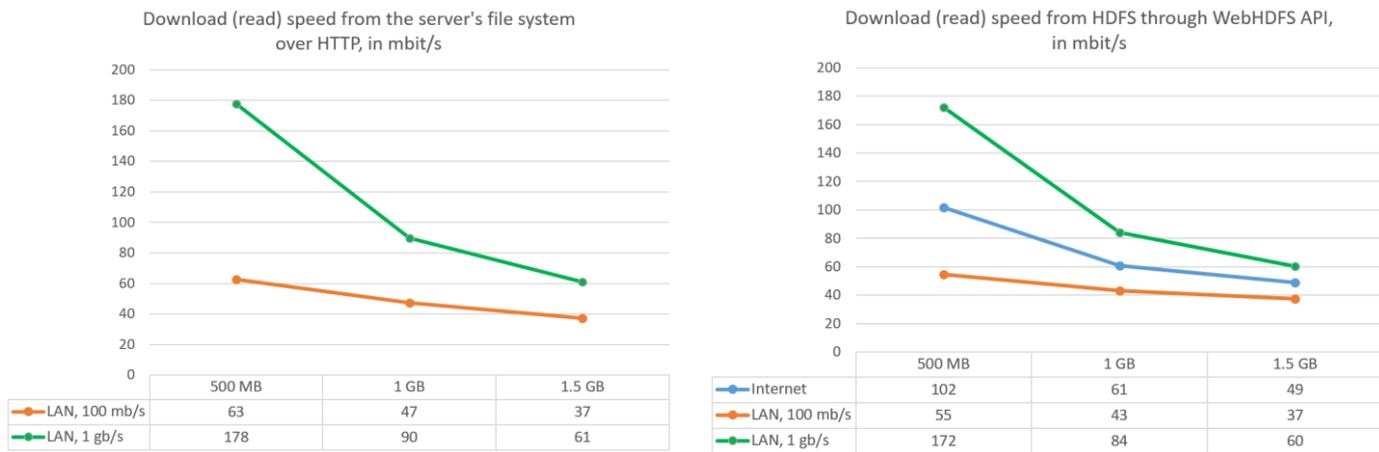

a) Download speed from the server's file system over HTTP    b) Download speed from HDFS through WebHDFS API

Fig. 7. Download (read) speed from the server's file system over HTTP
and from HDFS through WebHDFS API, for files from 500 to 1500 MB

It is easily noticeable that the lines of reading data from the server's file system over HTTP look quite similar to the lines of reading data from HDFS through the WebHDFS API. Especially on figure 7 (for large file sizes), lines are almost identical, so the absolute values of the read speed. It doesn't matter if files are read from the HDFS distributed file system via the WebHDFS API or from the HTTP server's file system. *If the reading is done using PHP's cURL library, the speed always decreases as the file size increases, and even at the same rate*. This proves that the fault for the decreasing read speed is not in the WebHDFS API, but in the php's cURL library. It should be stated here that it applies to the php's cURL library only, not to the entire cURL project. The WebHDFS API was also tested by stand-alone command-line cURL tools (on both Windows and MacOS) and they achieve constant download speed for all file sizes.

**CONCLUSION**

After performing dozens of experiments, it could be concluded that:

1. WebHDFS API allows data exchange with the Hadoop Distributed File System (HDFS) at very high speeds, and in general it is not the limitation factor, but the speed of the network itself.

2. The speed of writing files to the HTTP server's file system, and also to the distributed HDFS, through the WebHDFS API, by using cURL library for php does not depend on the files size, but remains constant and is limited only by the network capacity.

3. The speed of reading files from the HTTP server's file system, and also from the distributed HDFS, through the WebHDFS API, by using cURL library decreases rapidly as the file size increases.

4. The reason for the decreasing read (download) speed is not the server side itself, but the implementation of the cURL library, distributed together with the PHP interpreter.

5. When reading files from the server's file system over HTTP or from WebHDFS API by using PHP and cURL, it is mandatory that the PHP interpreter is configured to use a larger amount of RAM memory than the size of the files being read. This is expected since the data transfer happens in multiple small network packets, but in order to reconstruct the file from them, they must be stored and arranged in a common buffer (located within the RAM memory).

ACKNOWLEDGEMENTS

This paper is supported by project 23-FEEA-01 "Development of models and simulations with different application areas", funded by the Research Fund of the "Angel Kanchev" University of Ruse.